# Towards Evaluating Large Language Models for Graph Query Generation

Siraj Munir[0000−0003−0423−7585] and Alessandro Aldini[0000−0002−7250−5011]

Department of Pure and Applied Sciences, University of Urbino Carlo Bo, Urbino, Italy
s.munir@campus.uniurb.it, alessandro.aldini@uniurb.it

Abstract. Large Language Models (LLMs) are revolutionizing the landscape of Generative Artificial Intelligence (GenAI), with innovative LLMbacked solutions emerging rapidly. However, when applied to database technologies, specifically query generation for graph databases and Knowledge Graphs (KGs), LLMs still face significant challenges. While research on LLM-driven query generation for Structured Query Language (SQL) exists, similar systems for graph databases remain underdeveloped. This paper presents a comparative study addressing the challenge of generating Cypher queries – a powerful language for interacting with graph databases – using open-access LLMs. We rigorously evaluate several LLM agents (OpenAI ChatGPT 4o, Claude Sonnet 3.5, Google Gemini Pro 1.5, and a locally deployed Llama 3.1 8B) using a designed few-shot learning prompt and Retrieval Augmented Generation (RAG) backed by Chain-of-Thoughts (CoT) reasoning. Our empirical analysis of query generation accuracy reveals that Claude Sonnet 3.5 outperforms its counterparts in this specific domain. Further, we highlight promising future research directions to address the identified limitations and advance LLMdriven query generation for graph databases.

## 1 Introduction

Large Language Models (LLMs) have become central to numerous Artificial Intelligence (AI) driven applications, ranging from chat agents to Retrieval Augmented Generation (RAG) systems. They have significantly advanced the AI landscape by offering robust solutions to complex tasks. However, certain research and application areas still pose challenges to LLMs, demanding further investigation and development [1–3]. Text2Query generation stands out as one such area that has garnered considerable attention within the research community.

Recent literature and developer libraries highlight the potential of LLMs for automating and enhancing query generation. However, current efforts are predominantly focused on Structured Query Language (SQL), the classic query language for relational databases. This bias towards SQL can be attributed to several factors:

- SQL is a robust, well-established standard that offers a comprehensive set of features. It has served as the foundation for many subsequent query languages.
- SQL is ISO certified, ensuring a high degree of structure and standardization across implementations.
- Thousands of companies rely on SQL for their database management needs, reflecting its widespread adoption and maturity.

While SQL remains a powerful and versatile language, alternative query languages offer distinct advantages for specific domains. Graph databases, for instance, expose limitations in SQL's ability to efficiently handle complex relationships and interconnected data. Cypher[1], a query language introduced by Neo4j[2], excels in handling the graph data paradigm. As the information landscape becomes increasingly interconnected, traditional approaches are being optimized or superseded by more specialized solutions. Graph Query Language (GQL), an ISO-standardized format for interacting with graph data sources, exemplifies this trend.

This paper focuses on comparing the generative capabilities of existing open LLMs specifically for the Cypher query language. By "open", we refer to LLMs that are either open-source or provide a free chat interface. For this study, we selected OpenAI ChatGPT 4o, Claude Sonnet 3.5, Google Gemini Pro 1.5, and a locally deployed Llama 3.1 8B Chat variant. Our focus on Cypher stems from Neo4j's powerful capabilities and robust support for modeling and maintaining Knowledge Graphs (KGs). KGs are a proven and robust semantic data structure for handling diverse data represented as graphs, making Cypher a highly relevant target for evaluating LLM capabilities in this context. This comparative analysis investigates three key aspects:

- Cypher Syntax Generation: Assessing the LLMs' ability to generate syntactically correct Cypher queries based on given prompts.
- Factual Understanding: Evaluating the LLMs' comprehension of factual information provided within the prompts and their ability to translate this information into accurate queries.
- Chain of Thought (CoT) Reasoning: Examining the LLMs' capacity to perform multi-step reasoning and incorporate logical deductions into the generated Cypher queries.

The key contributions of this study are:

---

[1] https://neo4j.com/docs/cypher-manual/current/introduction/
[2] https://neo4j.com/

- To evaluate and validate the Cypher syntax generation capabilities of openaccess LLMs.
- To assess the factual understanding of these LLMs based on the information provided in prompts.
- To evaluate the Chain of Thought (CoT) reasoning capabilities of these models in the context of Cypher query generation.

The remainder of this paper is structured as follows. Section 2 provides a brief overview of the background and related technologies relevant to our work. Section 3 presents a systematic review of related research, highlighting recent advancements in LLM-driven query formulation. Section 4 details our proposed methodology, including the workflow, dataset, and evaluation criteria. Section 5 presents and discusses the results of our comparative analysis. Finally, Section 6 concludes the paper and outlines promising directions for future research.

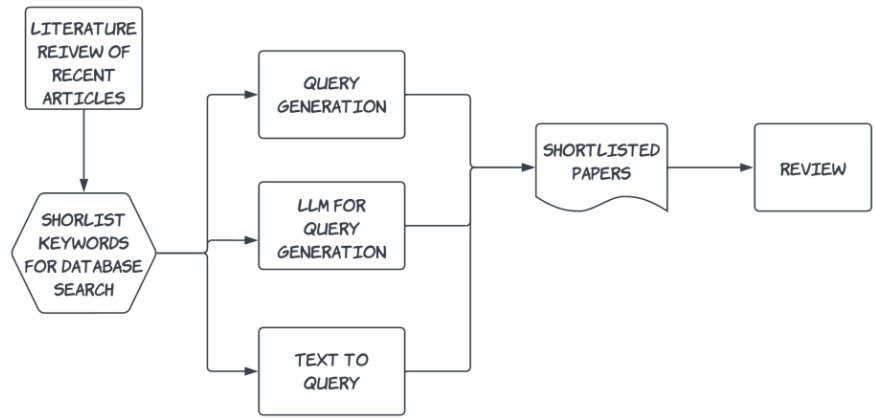

Fig.1. Literature Review Pipeline

## 2    Background and related technologies

This section sheds light on the formal definitions and technologies we opted for this work.

### 2.1   Natural Language Processing and Natural Language Understanding

Natural Language Processing (NLP) is a branch of computational linguistics that allows us to train machines to understand and interact with users using natural language [4]. After the latest advancements in AI, NLP has grown from simple grammar (rule-based) to deterministic probability-based solutions and more.

Natural Language Understanding (NLU) is a sub-area or branch of NLP that focuses on semantics and intent of input for contextual learning [5].

### 2.2   Generative Artificial Intelligence and Large Language Models

Generative AI (GenAI) is the new paradigm for complex user queries. GenAI models are more powerful and responsive than the preceding natural and visionbased models. They are autonomous in generating multimodal genres of outputs, e.g. text, text-text, text-image, image-text, text-video, etc.

LLMs are the foundational building block of GenAI applications. They provide powerhouse, memory, and reasoning capabilities to these applications.[3] Among all generative tasks, LLMs are widely used for text generation. There could be multiple reasons for it. For example, the availability of huge corpora, user demands, etc.

LLM Driven Agents There are some misconceptions when thinking of LLMdriven agents. LLM is like a brain for an agent. Whereas agents are far more complex than LLMs. Agents or intelligent agents are expected to be responsive either at the system or instruction level. As they perceive and react based on the environment they exist [6]. Now environment definition varies as per application area. For example, a robotic agent environment can be a house, a terrain, or a lab, a software agent's environment is the abstract access to computational resources, etc. LLM-driven agents use LLM as a decision factory for handling information flow across applications.[4] Formally, an agent is represented as $A$ and has several states or environmental conditions $S$ on which it acts making an action or decisions $a$.

$$A = (S, a) \qquad (1)$$

## 3   State of the Art

This section presents a systematic review of recent advancements in query formulation and information retrieval leveraging Large Language Models (LLMs). We focused on publications from top-tier venues such as ACM, IEEE, and ScienceDirect, employing a structured pipeline depicted in Figure 1 to ensure a comprehensive and rigorous literature review. Our survey reveals several key trends and contributions:

---

[3] https://www.ibm.com/topics/large-language-models
[4] https://blog.langchain.dev/what-is-an-agent/

### 3.1 LLM-Driven Query Generation

Several studies highlight the potential of LLMs to automate and enhance query formulation: Synthetic Query Generation: The study presented in [7] demonstrates the feasibility of using LLMs to generate synthetic queries for virtual assistants, achieving promising results by aligning generated queries with user intent. Query Optimization for Language Models: The author of [8] presented a preprint addressing the challenge of query formulation. Authors suggested that better-formulated queries lead to better information retrieval for language models. Document Generation from Queries: [9] presented an interesting study for generating documents from natural language queries via LLM. The authors evaluated baseline models and attained reasonable results. SQL Query Rewriting: Authors of [10] put forth a query re-writing model driven by LLM. The proposed approach takes SQL query as input and tries to suggest rewriting and optimizing query processing time. To validate, the authors thoroughly experimented with different LLM models and evaluated the proposed approach semantically and syntactically.

### 3.2 Contextualized Query Formulation

Recognizing the importance of user context, recent work has focused on incorporating contextual information into query generation: Context-Aware Query Reformulation: The study presented in [11] puts forth contextual-driven query reformulation approaches. Authors commented that former approaches did not consider user context while generating a query, hence they lacked in fidelity of query. The proposed approach incorporates user context by indulging user-aware generation.

### 3.3 Novel Query Languages and Paradigms

LLMs have also enabled the development of innovative query languages and paradigms: LLM-Based Programming with Query Languages: Authors in [12] highlighted a novel paradigm prompting in programming using LLM-based query language. The presented query language syntax is inspired by SQL query language. The authors also evaluated the presented approach on various query generation and reasoning tasks.

### 3.4 Beyond Traditional Retrieval: Generation to Retrieval

Challenging traditional retrieval approaches, some studies leverage LLMs for generation-to-retrieval methods: Clustering for Text Generation: The work in [13] proposed an interesting clustering approach for text generation. The presented model adopted a reverse approach, i.e., instead of retrieval to generation, they opted for the contextual generation to retrieval. Pilots show that the presented model outperformed state-of-the-art retrieval models on the TriviaQA and WebQ datasets.

This survey underscores the significant role LLMs play in advancing query formulation and information retrieval. From automating query generation to incorporating context and exploring novel paradigms, LLMs are transforming how we interact with and access information.

### 3.5 A Comparative Analysis

The surveyed literature highlights the rapid advancement of LLM-driven query formulation and information retrieval. To better understand the landscape of these advancements, we present a comparative analysis of the different approaches identified in our state-of-the-art review. This analysis focuses on key aspects such as query language targeted, context awareness, novelty of the approach, and evaluation methodology. We also include our proposed approach in this analysis, highlighting its unique contribution and positioning within the existing research.

Table 1. Comparative Analysis of LLM-Driven Query Formulation Approaches

| Approach | Query Language | Context Awareness | Novelty | Evaluation | Strengths | Weaknesses |
| --- | --- | --- | --- | --- | --- | --- |
| [7] | Not Specified | Limited | Moderate | User Intent Alignment | Aligns well with user intent | Limited context awareness |
| [8] | Not Specified | Not Addressed | Moderate | Information Retrieval Effectiveness | Improves retrieval accuracy | Evaluation focuses solely on retrieval |
| [9] | Natural Language | Limited | Moderate | Text Generation Quality | Generates coherent documents | Limited exploration of diverse query types |
| [10] | SQL | Limited | High | Semantic and Syntactic Correctness | Improves query efficiency | Focuses solely on SQL rewriting |
| [11] | Not Specified | High | High | Query Fidelity and Relevance | Incorporates user context effectively | Lacks detailed evaluation on diverse datasets |
| [12] | SQL-inspired | Moderate | High | Query Generation and Reasoning Tasks | Novel programming paradigm | Limited real-world application demonstration |
| [13] | Not Applicable (Generation to Retrieval) | Moderate | High | Performance on TriviaQA and WebQ datasets | Outperforms traditional retrieval models | Limited generalizability to other domains |
| Ours | Cypher | Moderate | High | Syntax, Factual Accuracy, CoT Reasoning | Focuses on the emerging Cypher language and KG domain; Evaluates CoT reasoning capabilities for complex query generation. | Limited to freely available LLMs; Evaluation dataset may not cover all Cypher functionalities. |

This comparative analysis reveals that while existing approaches demonstrate the potential of LLMs in query formulation, they often overlook the specific challenges and opportunities presented by graph databases and query languages like Cypher. Our proposed approach directly addresses this gap by evaluating and comparing the capabilities of open LLMs for Cypher query generation, contributing valuable insights to this emerging research area.

## 4  Methodology

This section will highlight the workflow we opted to evaluate CoT and user prompts. The workflow is presented in Figure 2. For each LLM, we followed the same workflow mentioned earlier. Let us dive deep into more details in a phase-wise manner.

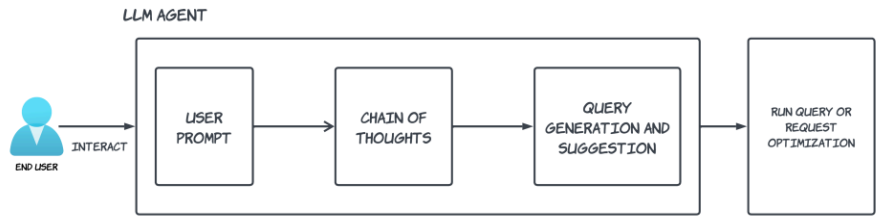

Fig.2. LLM driven Workflow for Query Generation

### 4.1  Workflow for LLM driven Query Generation

In this section, we will expand each phase presented in Figure 2. Starting with user interaction, each LLM model or agent takes a user query and prompt in natural text and then, based on the given CoT, tries to make up a query satisfying the user criteria, or else the user can ask for further optimizations. But note that for evaluation we considered the first response and measured accuracy and consistency. Now let us comment on each phase individually.

User Prompt For modeling user prompts, we opted few-shot learning strategy where we gave example queries along with a possible Cypher query. For all the mentioned examples we targeted the smart or intelligent workspace use case. For example, consider a simple user query: How to find a relationship between Person Siraj Munir and Organization Smart Industry. Listing 1.1. Prompt Engineering

 Analyze the following question and provide insights:
          Question: "{user-question}"
             1.   Identify the main entities mentioned (e.g., personnames, locations, roles).

   2. Determine the type of information being requested(e.g., personal details, relationships, paths).
   3. Identify any constraints or specific conditionsmentioned.
   4. Suggest potential follow-up questions or areas forclarification.

Task: Provide a comprehensive answer to the question using the available information. Follow these steps:
   1. Summarize the key points from the questionanalysis.
   2. Identify the most relevant pieces of informationfrom the database results.
   3. Draw connections between the question, analysis,and available data.
   4. Formulate a clear and concise answer to theoriginal question.
   5. If there are any limitations or uncertainties inthe data, mention them.
   6. Suggest potential follow-up questions or areas forfurther investigation.
   7. Structure your response to show your reasoningprocess

This was the general pattern we gave for each query. Then following this we added a few-shot example to give the LLM agent a bit of context for better generation of queries.

Listing 1.2. Few-Shot Examples

1. Find person by name: MATCH (p:Person {name: 'Person Name'}) RETURN p
2. Find all persons: MATCH (p:Person) RETURN p
3. Find persons with similar names: MATCH (p:Person) WHERE p.name =~ '(?i).*Name.*' RETURN p
4. Find connections: MATCH (p1:Person)-[r]-(p2:Person
   ) WHERE p1.name = 'Person1' RETURN p1, r, p2
5. To find connections between person and a role: MATCH p= shortestPath((p:Person {name: 'Person
   Name'})-[*..5]-(r:Role)) RETURN p
6. Find connections between person and a location:
   MATCH p= shortestPath((p:Person {name: 'Person
   Name'})-[*..5]-(l:Location)) RETURN p
7. Find connections between person and organization: MATCH p= shortestPath((p:Person {name: 'Person
   Name'})-[*..5]-(o:Organization)) RETURN p

Chain of Thoughts For modeling CoT, we defined a comprehensive pipeline so that the agent self-assists its response before the generation of any text. Consider the following CoT example for the smart workspace.

Listing 1.3. Chain of Thoughts Example

Chain-of-Thoughts:
   Question: "How to find relationship between Person Siraj Munir and Organization Smart Industry" Analysis:
   Analyze the following question and provide insights:
   Question: "How to find relationship between Person Siraj Munir and Organization
      Smart Industry"

1. Identify the main entities mentioned (e.g., personnames, locations, roles).
2. Determine the type of information being requested(e.g., personal details, relationships, paths).
3. Identify any constraints or specific conditionsmentioned.
4. Suggest potential follow-up questions or areas forclarification.

Please provide your analysis in a structured format.
**Main Entities:**
* Person: Siraj Munir
* Organization: Smart Industry
**Type of Information:**
* Relationship between Person and Organization
**Constraints/Conditions:**
* None mentioned
**Potential Follow-up Questions:**
* What is the nature of the relationship between Siraj Munirand Smart Industry (e.g., employee, contractor, supplier)
      ?
* Is Siraj Munir a current or former employee of Smart
      Industry?
* What is Siraj Munir's role or position within SmartIndustry, if any?
* Are there any specific projects or initiatives that SirajMunir has worked on with Smart Industry?
* Can we find any publicly available information about Siraj
      Munir's relationship with Smart Industry? (e.g., LinkedIn profile, news articles)

Using this mentioned CoT we were able to enforce the agent to generate specified Cypher syntax.

Query Generation For query generation, we simply asked the agent to follow the preceding flow and generate a single query for each user query. Example query generation is shown below.

Listing 1.4. Query Generation Example

Cypher query to find the relationship between Siraj Munir and Smart Industry:

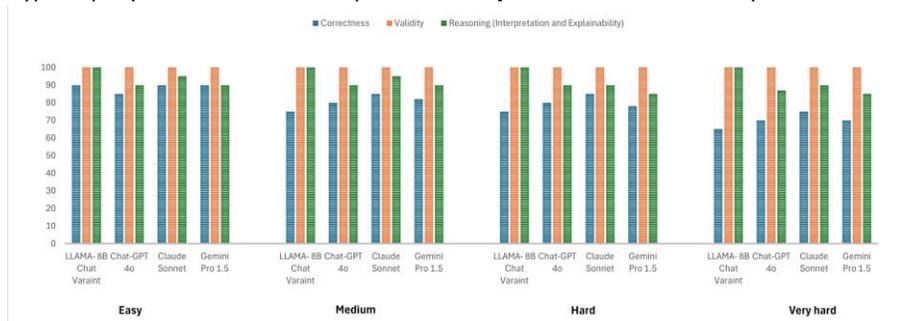

Fig.3. Comparison of LLM Agents

```
MATCH (p:Person {name: 'Siraj Munir'})-[:WORKS_FOR|
    FOLLOWED_BY|MENTIONS|LOCATED_AT|ASSOCIATED_WITH]-(
    o:Organization {name: 'Smart Industry'}) RETURN p, r, o
```
This query will return all relationships between Siraj Munir and Smart Industry, including WORKS_FOR, FOLLOWED_BY, MENTIONS, LOCATED_AT, and ASSOCIATED_WITH. If no relationships are found, the query will return an empty result set.

Note: This query assumes that the relationships between Person and Organization are defined in the graph database . If the relationships are not defined, the query will not return any results.

To get more specific results, you can add more constraints to the query, such as:
```
MATCH (p:Person {name: 'Siraj Munir'})-[:WORKS_FOR]->(o
    :Organization {name: 'Smart Industry'}) RETURN p, r
    , o
```
This query will return only the WORKS_FOR relationship between Siraj Munir and Smart Industry.

To get the shortest path between Siraj Munir and Smart Industry, you can use the shortestPath function:
```
MATCH p= shortestPath((p:Person {name: 'Siraj Munir'})
    -[*..5]-(o:Organization {name: 'Smart Industry'})) RETURN p
```
This query will return the shortest path between Siraj Munir and Smart Industry, including all intermediate nodes.

Note: The shortestPath function returns a path, not a relationship. To get the relationships along the path, you can use the unwind function.

Notice that we included the output response of the Llama-8B LLM agent here. The first version of the query is generated based on the user question and then improvised later using CoT. Another important aspect is that we only considered the agent's first output. Whereas in general practice end-users can ask for further clarification or even regeneration of query. The subsequent section will discuss the validation pipeline we designed to evaluate the LLM agents.

### 4.2 Validation and Evaluation

Validation is paramount for establishing the boundaries and capabilities of a proposed approach. To validate and evaluate our workflow, we created a synthetic dataset, described in a subsequent subsection. This dataset served as the basis for comparing the performance of each LLM model across different query complexity levels.

We defined four levels of query generation tasks: easy, medium, hard, and very hard. For each task, we compared the LLM-generated query against a "ground truth" query, evaluating its accuracy and consistency. The accuracy is mathematically expressed as:

$$\text{Accuracy} = \frac{\text{Number of Correct Queries}}{\text{Total Number of Queries}} \times 100 \quad (2)$$

Similarly, the error rate can be evaluated as:

$$\text{Error Rate} = \frac{\text{Number of Incorrect Queries}}{\text{Total Number of Queries}} \times 100 \quad (3)$$

We will delve into a detailed, section-wise discussion of our evaluation methodology and results to ensure clarity and readability.

### 4.3 Dataset

A robust and representative dataset is crucial for the validation and evaluation of any LLM-driven system. Acknowledging this, and drawing upon our understanding of Cypher query language syntax, we generated a synthetic dataset consisting of 100 queries related to the "smart workspace" domain. For generating these queries, we employed the Claude Sonnet LLM model, given its proven track record in code generation tasks.[5] It is important to note that we exclusively utilized the free, publicly accessible version of Claude Sonnet for this purpose. Table 2 provides a breakdown of the query distribution across different complexity levels.

Table 2. Dataset Summary

| Easy | 25 |
|---|---|
| Medium | 25 |
| Hard | 25 |
| Very Hard | 25 |

We leveraged this dataset to compare the output generated by each LLM agent. The following section will elaborate on the evaluation metrics employed in this study.

### 4.4 Performance Evaluation

Our evaluation of the presented workflow was based on the following criteria:

- Correctness: Measures the factual and syntactical accuracy of the LLMgenerated Cypher query. A query is deemed correct if it is both semantically and syntactically sound, aligning with the expected outcome based on the user query and prompt.
- Validity: Assesses the consistency and logical coherence of the LLM's response. A valid response demonstrates that the agent's intermediate reasoning steps are appropriate and lead to a well-formed query.

---

[5] https://www.vellum.ai/llm-leaderboard

- Reasoning (Interpretation and Explainability): Evaluates the clarity, coherence, and depth of the thought process and explanations provided by the LLM. A high reasoning score indicates that the agent can effectively interpret the user query, follow the CoT, and provide human-understandable justification for its generated query.

Based on these criteria, we observed that Claude Sonnet 3.5 yielded the most favorable results, followed by OpenAI ChatGPT 4o, Google Gemini Pro 1.5, and lastly, Llama 3.1 8B. The lower performance of Llama-8B can be attributed to our use of a smaller, locally deployed variant due to resource constraints. It is important to emphasize that this evaluation focuses solely on the free, publicly accessible versions of these models.

Figure 3 illustrates the @Hit 1 accuracy and error matrix-based performance of each LLM agent. This evaluation aimed to provide researchers and developers with insights into the capabilities and limitations of readily available, open-source LLM options. The subsequent section will delve into the strengths, weaknesses, and potential research directions associated with each model.

### 4.5 Discussion and Results

This section analyzes the performance of individual LLM models on our designed task, discussing their strengths, weaknesses, and potential avenues for improvement.

Performance Across Complexity Levels Easy Tasks: All models exhibited commendable performance on easy tasks, achieving near-perfect scores in Correctness, Validity, and Reasoning. Llama-8B, while showing slightly lower reasoning and correctness, still maintained a high level of performance. The nearidentical results of Chat-GPT 4.0, Claude Sonnet, and Google Gemini suggest that these models effectively handle straightforward queries.

Medium Tasks: As expected, the overall performance saw a slight dip in correctness as the task complexity increased to the medium level. Validity and reasoning scores, however, remained consistently high. Chat-GPT 4.0 and Claude Sonnet maintained strong performance across all categories, while Gemini and Llama-8B Chat exhibited more noticeable drops in correctness, indicating potential struggles with moderately complex queries.

Hard Tasks: Moving to hard tasks led to greater performance variations among the models, particularly in correctness and reasoning. Chat-GPT 4.0, while excelling in validity and reasoning, showed a marked decrease in correctness compared to the previous levels. Llama-8B faced the most significant challenges, with a clear decline in correctness and reasoning, highlighting its limitations in handling the increased complexity. Gemini performed marginally better than Llama-8B but still lagged behind the top performers. These findings suggest that as query complexity rises, the ability of LLMs to maintain logical reasoning and correctness becomes more variable.

Very Hard Tasks: The most demanding task level revealed significant performance degradation across all models, especially in correctness. Llama-8B exhibited the most substantial decline, particularly in its ability to generate accurate Cypher code. Interestingly, Chat-GPT 4.0, despite achieving the highest reasoning scores, experienced a significant drop in correctness, indicating challenges in translating its sound reasoning into accurate Cypher outputs. Claude Sonnet, while maintaining relatively good validity and reasoning scores, also showed reduced correctness. Gemini performed on par with Chat-GPT 4.0, experiencing similar struggles with maintaining correctness. These findings underscore the limitations of current LLMs in consistently achieving high correctness, even when demonstrating robust reasoning capabilities.

Analysis of Key Findings Our analysis reveals several key findings regarding the performance of open-access LLMs in generating Cypher queries:

- Correctness Declines with Complexity: There is a clear trend of decreasing correctness as task difficulty increases. Chat-GPT, Gemini, and Claude consistently outperform Llama-8B in more challenging scenarios, likely due to differences in training data size and model architecture.
- Reasoning Does Not Guarantee Correctness: Even models with demonstrably strong reasoning capabilities, such as Chat-GPT, struggle to produce consistently correct outputs for highly complex queries. This suggests that robust reasoning alone is insufficient for achieving high accuracy in Cypher query generation.
- High Validity Across Models: A positive observation is that validity scores remain relatively high across all models and task difficulties. This suggests that most LLMs can maintain logical consistency in their responses, even when struggling to achieve perfect correctness.
- Variations in Validity Decline: While most models maintain high validity, Llama-8B shows a more noticeable decline in validity as tasks become more challenging. This could point to limitations in the model's ability to handle complex relationships and constraints consistently.
- ChatGPT Excels in Reasoning and Explainability: Chat-GPT 4.0 demonstrates superior reasoning and explainability compared to its counterparts, especially as tasks become more intricate. Claude Sonnet and Gemini perform well but lag slightly behind Chat-GPT in this regard.
- Llama-8B's Reasoning Limitations: Llama-8B, while performing adequately on less demanding tasks, reveals a significant gap in reasoning ability as task complexity escalates. This highlights the challenges faced by smaller LLM variants in handling complex, multi-step reasoning effectively.

Summary of Results Our findings indicate that while all evaluated models perform well on simple Cypher query generation tasks, their capabilities diverge as complexity increases. Chat-GPT 4.0 consistently exhibits strong reasoning skills but still grapples with correctness in highly complex scenarios. Claude Sonnet and Gemini Pro 1.5 demonstrate strong validity and reasoning throughout but experience a similar decline in correctness for difficult queries. Llama-8B Chat Variant consistently lags behind on harder tasks, particularly in correctness and reasoning.

A key takeaway is that improvements in correctness, especially for complex queries, are crucial for the future development of LLMs for this domain. Ensuring that models can maintain both high reasoning ability and accuracy will be paramount for real-world applications that demand reliable and precise query generation.

## 5   Conclusion and Future Work

This study addressed the emerging and underexplored challenge of generating Cypher queries using open-access LLMs. Our findings demonstrate that while current LLMs show promise in understanding natural language and producing syntactically valid Cypher code, they struggle to achieve consistently high accuracy, especially for complex queries that demand sophisticated reasoning. Claude Sonnet 3.5 exhibited the most consistent performance across varying complexity levels, suggesting its potential for such tasks. However, as our analysis revealed, even the most capable models struggle to maintain high accuracy as query complexity increases, highlighting the need for further research and development in this area.

Our work paves the way for numerous promising future research avenues:

Specialized LLM Training: Training LLMs on a large and diverse corpus of Cypher queries and graph data could significantly enhance their performance. Focusing on domain-specific datasets relevant to use cases like knowledge graph construction and querying would further improve their accuracy and relevance.

Enhanced Prompt Engineering: As our study revealed limitations in handling complex queries, exploring advanced prompt engineering techniques, such as incorporating graph schema information or user interaction history, could provide the LLMs with richer context and lead to more accurate query generation.

Hybrid Approaches: Combining the strengths of LLMs with traditional symbolic AI techniques for query understanding and semantic parsing presents a promising direction. Hybrid systems could leverage the flexibility and learning capacity of LLMs while benefiting from the precision and transparency of symbolic methods.

Explainability and Verifiability: Developing methods to make the LLM's reasoning process more transparent and its generated queries easily verifiable by human users will be crucial. Enhanced explainability builds trust and facilitates debugging, while verifiability ensures that the generated queries align with user intent and domain-specific constraints.

The field of LLM-powered graph database interactions is still in its infancy. Continued research and development in these directions hold immense potential for transforming how users interact with and leverage the power of graph data, ultimately driving innovation across various domains reliant on knowledge discovery and complex relationship analysis.